\documentclass[prl,twocolumn]{revtex4-1}
\usepackage{amsmath}
\usepackage{amssymb}
\usepackage{array}
\usepackage{mathrsfs}
\usepackage{tabularx}
\usepackage{graphicx}
\usepackage{dcolumn}
\usepackage{bm}
\usepackage{graphicx}
\usepackage{color}
\usepackage{diagbox}

\begin{document}
\title{Experimental Three-State Measurement-Device-Independent Quantum Key Distribution with Uncharacterized Sources}
\author{Xing-Yu Zhou$^{1,2,3,\dagger}$}
\author{Hua-Jian Ding$^{1,2,3,\dagger}$}
\author{Chun-Hui Zhang$^{1,2,3}$}
\author{Jian Li $^{1,2,3}$}
\author{Chun-Mei Zhang$^{1,2,3}$}
\author{Qin Wang$^{1,2,3}$}
\email{qinw@njupt.edu.cn}

\affiliation{$^{1}$Institute of quantum information and technology, Nanjing University of Posts and Telecommunications, Nanjing 210003, China.}
\affiliation{$^{2}$"Broadband Wireless Communication and Sensor Network Technology" Key Lab of Ministry of Education, NUPT, Nanjing 210003, China.}
\affiliation{$^{3}$"Telecommunication and Networks" National Engineering Research Center, NUPT, Nanjing 210003, China.}

\begin{abstract}
The measurement-device-independent quantum key distribution (MDI-QKD) protocol plays an important role in quantum communications due to its high level of security and  practicability. It can be immune to all side-channel attacks directed on the detecting devices. However, the protocol still contains strict requirements during state preparation in most existing MDI-QKD schemes, e.g., perfect state preparation or perfectly characterized sources, which are very hard to realize in practice. In this letter, we investigate uncharacterized MDI-QKD by utilizing a three-state method, greatly reducing the finite-size effect. The only requirement for state preparation is that the state are prepared in a  bidimensional Hilbert space. Furthermore, a proof-of-principle demonstration over a 170 km transmission distance is achieved, representing the longest transmission distance  under the same security level on record. 

PACS number(s): 03.67.Dd, 03.67.Hk

\end{abstract}

\maketitle

\section{Introduction}
The quantum computer \cite{QC1,QC2}, which is capable of exponentially faster execution speeds, is threatening conventional public cryptosystems based on computational complexity. Fortunately, the quantum key  distribution (QKD) has come into being, which allows two remote legitimate users (Alice and Bob) to share cryptographic keys with information-theoretic security based on quantum physics \cite{HK,PShor,DMayers}, causing a new revolution in secure communication. Since the first BB84 protocol \cite{BB84} was put forward in 1984, plenty of works have been devoted to security both in theory and experiment \cite{AT1,AT2,AT3,AT4,DI}. Among  them, the device-independent (DI) QKD possesses the highest level of security. However, it is very sensitive to losses from either of the channels or detection equipment, and thus difficult to realize with current technology. Then, the measurement-device-independent QKD (MDI-QKD) was proposed \cite{Side,MDI}, removing all risk of side-channel attacks directed at the detector and reaching a good  balance between security and practicality. Up to today, many MDI-QKD experimental demonstrations have been performed \cite{EMDI1,EMDI2,EMDI3, EMDI4}, showing great potential for real-life implementations. 

Although the MDI-QKD protocol can defend against all side-channel attacks, it does still have requirements for the state-preparation process that can be a great challenge in practical implementations. To protect against state-preparation imperfections, some countermeasures have been raised, such as loss-tolerant methods \cite{LT,LTMDI}. Nonetheless, they require a  detailed characterization of the prepared states, making the QKD system much more complicated. Recently, Yin \emph{et al.} proposed a method \cite{Yin1,Yin2} incorporating mismatched-basis data  into the phase-error rate calculation. The only assumption is that the prepared states are in a two-dimensional Hilbert space, greatly reducing the complexity of experiment realization. Soon afterwards, some  related theoretical and experimental works \cite{Unch1,Unch2,Unch3,Unch4} were published. However, they are inefficient in either the phase-error estimations \cite{Yin2,Unch1,Unch2} or the decoy-state method employed \cite{Unch3}. Here we investigate the three-state MDI-QKD protocol with uncharacterized sources, and carry out a corresponding experimental demonstration. Under the same security level, we achieve a new record distance of 170 km with pigtailed fibers. 

Before presenting the protocol, we should declare that our protocol can be implemented with different decoy-state methods, e.g., a three-intensity \cite{XYZ} or four-intensity \cite{Wang1} scheme. Here for simplicity, we only use the three-intensity method for illustration. Now let's briefly introduce the scheme, and more detailed descriptions  will be available in the supplementary material  \cite{Supp}.  

First, Alice and Bob randomly prepare their weak coherent sources (WCS) into three different intensities $(u, v, o)$ in a Z or X basis, then they send the signal out to Charlie; here $u$ and $v$ each represents the signal and the decoy state, respectively; $o$ refers to the vacuum state. In the Z basis, the bidimensional encoding states prepared by Alice and Bob can be denoted as $\left| {{\varphi _m}} \right\rangle $ and $\left| {{\varphi '_n}} \right\rangle $ respectively, where $m,n \in \{ 0,1\} $. Noting that, here $\left| {{\varphi _m}} \right\rangle $ and $\left| {{\varphi '_n}} \right\rangle $ do not have to be ideal BB84 states. 
In the X basis, in contrast to the original uncharacterized protocols \cite{Yin2,Unch1,Unch2,Unch4}, here only one state is required. By  appending corresponding phases (${\theta }, {\theta '}$) to the superposition of the states in the Z basis, the final states can be written as
\begin{equation}\label{111}
\begin{array}{l}
\left| {{\varphi _2}} \right\rangle  = {c_{0}}\left| {{\varphi _0}} \right\rangle  + {c_{1}}{e^{i{\theta }}}\left| {{\varphi _1}} \right\rangle ,\\
\left| {{{\varphi '_2}}} \right\rangle  = {{c'_{0}}}\left| {{{\varphi '_0}}} \right\rangle  + {{c'_{1}}}{e^{i{{\theta '}}}}\left| {{{\varphi '_1}}} \right\rangle,
\end{array}
\end{equation}
where ${c_{x}}$ and ${{c'_{x}}}$ are non-negative real numbers ($x \in \{0, 1\}$). These uncertainty values indicate  that the states are also uncharacterized as is the case in the Z basis, and can even be unknown to  Alice and Bob.

Second, Charlie performs Bell-state measurements on the pulse-pairs received from Alice and Bob, and announces the results of successful events. Then Alice and Bob exchange the basis-choice  information. Differing from traditional MDI-QKD protocol, here they keep all effective events including those with matched or mismatched bases. Here, a matched (or mismatched) basis means that Alice and Bob chose the same (or different) basis.

Finally, Alice and Bob carry out parameter estimation and post-processing, obtaining the secure keys. In the process, the phase error rate ${e_p}$ is estimated by utilizing the data from matched and mismatched bases:
\begin{equation}\label{eppp}
{e_p} \le \frac{{2{p_{00}} + 2{p_{11}} + {p_{01}} + {p_{10}} + 2\sqrt {{p_{01}}{p_{10}}} f({c_0},{c_1},{{c'}_0},{{c'}_1})}}{{2({p_{00}} + {p_{11}} + {p_{01}} + {p_{10}})}}.
\end{equation}
For the convenience of the experiment,  we only consider the $\left| {{\Psi ^ - }} \right\rangle$ projection measurement. ${p_{mn}}$ represents the yield of a successful event when Allice (Bob) sends a single photon pulse in the state $\left| {{\varphi _m}} \right\rangle $ ($\left| {{\varphi '_n}} \right\rangle $), $m,n \in \{ 0, 1, 2\} $, and
\begin{align}\label{123}
&f({c_0},{c_1},{{c'_0}},{{c'_1}})  \nonumber \\
=& \frac{{{{\left| {\sqrt {{p_{22}}}  + \sqrt {{p_{00}}} {c_0}{{c'_1}} + \sqrt {{p_{11}}} {c_1}{c'_0}} \right|}^2} - {p_{01}}c_0^2{c'_0}^2 - {p_{10}}c_1^2{c'_1}^2}}{{2\sqrt {{p_{01}}{p_{10}}} {c_0}{c'_0}{c_1}{c'_1}}}.
\end{align}
For a better estimation of ${e_p}$, $f({c_0},{c_1},{{c'_0}},{{c'_1}})$ should be minimized according to several constraints (see supplementary material for more details \cite{Supp}).
 
Finally, the secure key rate is accordingly given by:
\begin{equation}\label{R}
R = {({p_u}{p_{Z|u}})^2}\left\{ {{{({e^{ - u}}u)}^2}{Y_{11}}[1 - H({e_p})]  - {Q_{uu}^{ZZ}}fH({E_{uu}^{ZZ}})} \right\},
\end{equation}
where $H(x) =  - x\log_2 (x) - (1 - x)\log_2 (1 - x)$ is the binary Shannon entropy function; $Y_{11}$ is the single-photon yield; $Q_{uu}^{ZZ}$ and  $E_{uu}^{ZZ}$  are gain and average quantum bit error rate (QBER), respectively, when Alice and Bob both choose a signal state with a Z basis;  $f$ is the error correction efficiency which we assign to be 1.16; ${p_u}$ is the probability of preparing signal states, and ${p_{Z|u}}$ is the probability of choosing the Z basis conditional on the signal state.
 Note that, the X basis is only for error estimation and the Z basis is for final key generation. Therefore, the three-state method in this work will not introduce an extra security problem.

\begin{figure*}[htb]
\centering
\includegraphics[width=16cm]{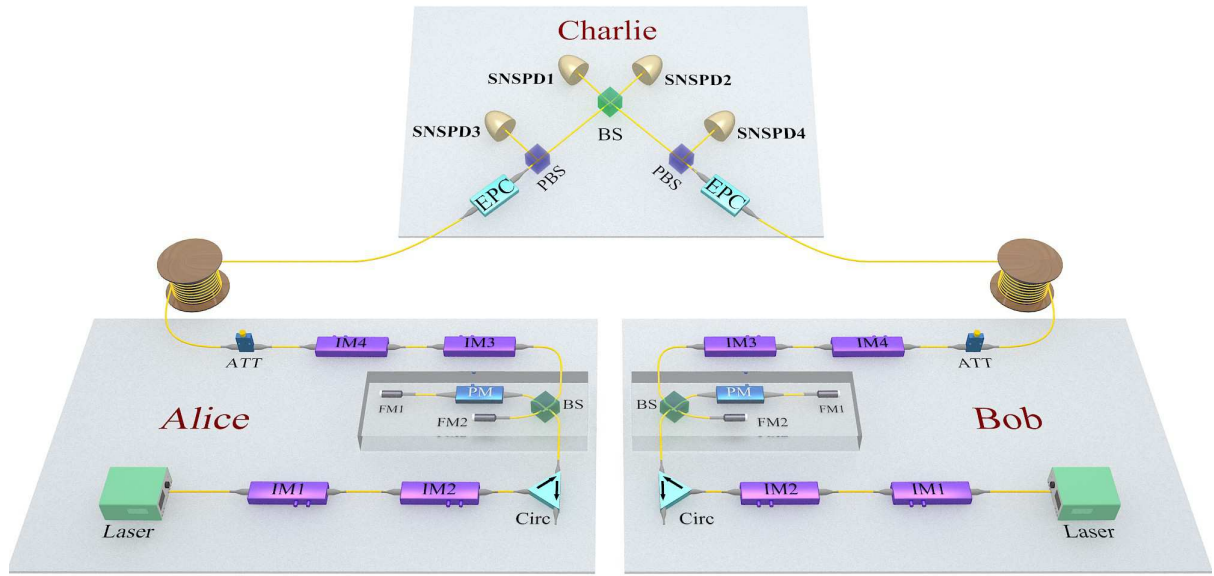}
\caption{Schematic setup of MDI-QKD with uncharacterized sources. Laser, continuous-wave laser; IM, intensity modulator; PM, phase modulator;  Circ, circulator; FM, Faraday mirror; ATT, attenuator; EPC, electronic polarization controller; BS: beam splitter; SNSPD, super-conducting nanowire single-photon detector.
}
\label{F1}
\end{figure*}

Our experimental setup is depicted in Fig. \ref{F1}. We apply the time-bin phase encoding scheme, and utilize intensity modulators (IMs) and  Faraday–Michelson interferometers (FMIs) \cite{FMI} as the key apparatuses of encoding. 
Two narrow linewidth continuous-wave lasers are used as light sources (Clarity NLL-1550-LP) whose frequencies are locked to the molecular absorption line with a center wavelength of 1550.51 nm. On either Alice or Bob's side, the light source is chopped by IM1 and IM2  into a pulse train with a 3 ns temporal width and a repetition rate of 50 MHz. Note that IM1 is used for decoy-state modulation and IM2 for extinction ratio improvement. 

On each side, the FMI, composed of a phase modulator (PM), a 50/50 beam splitter (BS) and two Faraday  mirrors (FMs),  separates each incoming pulse into two time bins with a temporal difference of 10.3 ns. The FM contains a ${45^{\rm{^\circ }}}$  Faraday rotator and a total reflector. Phase encoding in the X basis can be performed by controlling the PM to add an extra phase at the long arm. As demonstrated in Ref. \cite{FMI}, an FMI is insensitive to polarization disturbance in both arms, and thus can maintain stable performance in real-life applications. A circulator is inserted before the FMI to filter the laser light reflected backwards by the FMs.

The following IM3 and IM4 are applied for basis choice and time-bin encoding.  For the Z basis, only the former or the later pulse, each representing a 0 or 1 bit, is allowed to pass through. Here, two cascaded IMs are utilized to suppress the noise and get a high extinction ratio for either the vacuum state or the coding Z states. Benefiting from this, here the corresponding inherent error rate in the Z basis is only $0.15\%$. Moreover, IM3 and IM4 can also be used to balance the intensity difference in the two time slots of the X basis, caused by the insertion loss of the PM. The bits are encoded in the relative phase of the two pulses.  Additionally, normalization of the average photon number in  the two bases is performed. 
An arbitrary waveform generator controls the full encoding process at a clock rate of 50 MHz.
Then, the coded pulses are adjusted by an attenuator (ATT) to the single-photon level before being sent out to Charlie through quantum channels.

At the detection stage (Charlie),  four commercial super-conducting nanowire single-photon detectors (SNSPDs: TCOPRS-CCR-SW-85, SCONTEL company) work at 2.3 K, providing an $85\%$ detection efficiency at a dark count rate of 12 Hz after gating with a time window of 2 ns.
Charlie  performs the $\left| {{\Psi ^ - }} \right\rangle$ projection measurement by causing the incoming pulses to interfere. A time-to-digital converter records the clicks from SNSPD1 and SNSPD2, and counts the corresponding two-fold coincidences. 

Hong–Ou–Mandel (HOM) interference is the key point of MDI-QKD.  For better  interference, we implement a narrow linewidth continuous-wave laser with high accuracy in the frequency domain, which contributes to the easy overlap of two independent pulses temporally and spectrally.  Polarization correction modules are also applied by monitoring the reflection parts from the polarization beam splitters (PBSs) with SNSPD3 and SNSPD4 while the electronic polarization controllers (EPCs)  compensate for the polarization drift every 20 minutes.  
Owing to the stable HOM interference, the misalignment error rate in the X basis is $1.5\%$.   To achieve a free-running system, we compensate the phase drift  of the FMIs and calibrate the  direct-current bias voltages of the IMs by quickly scanning the voltages and analyzing the corresponding count events.  Considering the extra time consumed, the duty cycle of transmission is around 90$\%$, which can be further improved by machine learning techniques \cite{ML}. 

We run our experiment with two coiled commercial standard single-mode fibers (0.18 dB/km) with 100 km and 170 km transmission distance.
The overall efficiency including the EPC, PBS, BS and detectors at Charlie's side is $ 67\% $. In this work, we consider the finite-data size effect \cite{Unch1,Unch4} which is absent in Ref. \cite{Unch2}.  The total number of pulses sent by Alice or Bob is ${10^{13}}$.  All the intensities and probabilities are optimized for better performance (see supplementary material for details \cite{Supp}) according to the system parameters. After collecting all the data, we estimate the phase errors and calculate the final secure key rates.

\begin{figure}[htbp]
\begin{center}
\includegraphics[scale=0.32]{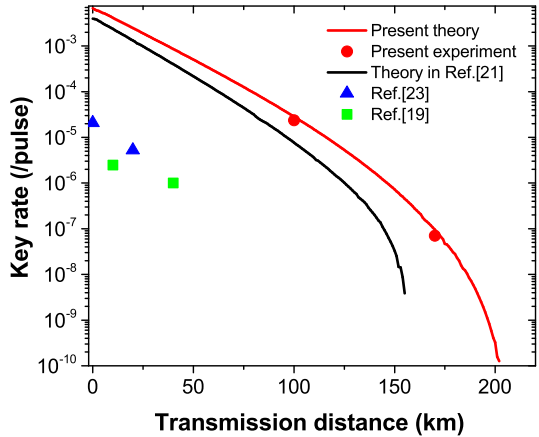}
\end{center}
\caption{Theoretical and experimental results of the key rate's dependence on the transmission distance. The red solid line is our simulation result; the dots represent our key rates with 100 km and 170 km fibers; the blue triangles are for the results in the work of Wang $et. al$ \cite{Unch2}, and the black solid line represents its theoretical result \cite{Yin2} using our experimental parameters; the green squares are the results of Tang $et. al$ \cite{LTMDI}. }
\label{F2}
\end{figure}

The theoretical and experimental results are shown in Fig. \ref{F2}. The transmission distance of our new method extends up to 200 km in theory. Compared with the result of Ref. \cite{Yin2} using the same experimental parameters, our result exhibits significant improvement both in transmission distance and key rate.   Thanks to the stability of the system, our experimental data show good agreement with the theoretical predictions.   At 100 km and 170 km, we obtain the key rates of  $2.36\times {10^{ - 5}}$ and $7.06\times {10^{ - 8}}$ per pulse respectively, and other crucial values for calculating the key rate are listed out in Table. \ref{T2}. Generally, the QBER is considerably low, which is one of the most decisive factors determining the key rate. 

In Fig. \ref{F2}, we also mark the experimental data of  Refs. \cite{LTMDI,Unch2} for  comparison. 
 Note that, the work of Wang $et$ $al.$ \cite{Unch2} presents an MDI-QKD with uncharacterized sources applying an original four-state method \cite{Yin2}, and the work of Tang $et$ $al.$ \cite{LTMDI} utilizes a loss-tolerant method. These experiments, including our work, possess similar security levels (except that \cite{Unch2} did not take the finite size effect into account), since they can not only  close the loopholes at the detection side, but also tolerate errors in state preparation. 
 By comparison, our result extends the secure transmission distance up to 170 km which is the longest demonstrated transmission distance under the same security level to date. In addition, although statistical fluctuation is considered, our key rate over short distances remains more than two orders of magnitude higher. Three main advantages contribute to our better performance: first, we adopt the three-state method and place a tight bound on phase error; second, the SNSPDs provide higher detection efficiency while maintaining a low dark-count rate; finally, cascade connection of IM3 and IM4 contributes to a lower error rate in the Z basis. 
Additionally, in the work of Tang $et.al$, state tomography is performed to characterize the state-preparation errors which increases complexity of realization. By contrast, no extra operation is taken in our work. We simply consider the mismatched-basis data, discarded in original MDI-QKD protocol, in the calculation. Moreover, one less state is prepared and the system can be further simplified. Especially, in some chip-based QKD system\cite{Chip1}, it may be difficult to prepare four states, and hence the method in this work will be more easily applicable.

\begin{center}
\begin{table}[thbp]
  \caption{Crucial values in the key rate formula: estimated single photon yield (${Y_{11}}$) and phase error rate (${e_p}$); measured values of QBER ($E_{uu}^{ZZ}$) and gain ($Q_{uu}^{ZZ}$) when Alice and Bob both prepare the signal state in the Z basis. ).
}
\renewcommand{\arraystretch}{1.3}
\begin{tabularx}{\linewidth}{XXXXXX}  \hline \hline
 Distance &${Y_{11}}$&${e_p}$&$E_{uu}^{ZZ}$&$Q_{uu}^{ZZ}$\\ \hline
100km &0.015&0.15&0.0017&$5.85 \times {10^{ - 4}}$\\ 
 170km &$7.68 \times {10^{ - 5}}$&0.27&0.0021&$7.53 \times {10^{ - 6}}$\\ \hline\hline
  \end{tabularx}
 \label{T2}
\end{table}
\end{center}

In summary, we have proposed a three-state MDI-QKD protocol with uncharacterized sources and carried out a corresponding experimental demonstration. In the protocol, one does not need to characterize the sources and thus the requirements of the state preparations are relaxed; meanwhile, it is immune to all side-channel attacks on the detection side due to its MDI characteristics. It thus possesses the highest level of security among all practical QKD protocols. Moreover, not only an improved phase estimation method but also a simple three-state scheme is implemented to obtain substantially enhanced performance compared with other uncharacterized or loss tolerant MDI-QKD schemes. Furthermore, by incorporating a state-of-art experimental setup, we have reached a record 170 km transmission distance under the same security level. Therefore, our work represents a further step towards the practical implementation of QKD.

We gratefully appreciate enlightened discussions from Y. G. J. Fan, S. Wang, H. Zhang and J. R. Zhu. This work is supported by the National Key R$\&$D Program of China (Grant Nos. 2018YFA0306400, 2017YFA0304100), the National Natural Science Foundation of China (Grants Nos. 61590932, 11774180, 61705110), the China Postdoctoral Science Foundation (Grant Nos. 2018M642281, 2019T120446), the Jiangsu Planned Projects for Postdoctoral Research Funds(Grant No. 2018K185C), and the Postgraduate Research and Practice Innovation Program of Jiangsu Province (Grant No. KYCX17$\_$0791).

$\dagger$ The two authors contribute equally to this work.

\end{document}